# Importance of phonon contribution to the free migration energy of the O vacancy in BaZrO$_3$

N. Raja[1], D. Murali[2], M. Posselt[3] and S. V. M. Satyanarayana[1, a)]

[1]*Department of Physics, Pondicherry University, Puducherry 605 014, India*

[2] *Department of Physics, Indian Institute of Technology Madras, Madras 600 036, India*

[3]*Helmholtz-Zentrum Dresden-Rossendorf, Institute of Ion Beam Physics and Materials Research, Bautzner Landstraße 400, 01328 Dresden, Germany*

[a)]*Electronic mail: svmsatya@gmail.com*

## Abstract

BaZrO$_3$ exhibits excellent proton conductivity and good high-temperature stability. It is therefore a promising electrolyte material for solid oxide fuel cells. The stability of BaZrO$_3$ at high temperatures is generally explained by the low diffusivity of the O vacancy. Present first-principle Density-Functional-Theory calculations show that the slow migration of the doubly positively charged O vacancy at high temperature cannot be solely caused by the ground-state migration energy but by the contribution of phonon excitations to the free migration energy. With increasing temperature, the effective barrier for oxygen vacancy migration increases. At about 1000K, which is the operating temperature of fuel cells, the calculated O vacancy diffusivity is more than one order of magnitude lower than that determined using solely the ground-state migration barrier. The calculated diffusivity data agree well with experimental results from literature. The present work reveals that the high-temperature stability of BaZrO$_3$ is mainly due to the phonon contribution to the free migration energy of the O vacancy.

BaZrO$_3$ (BZO), a cubic perovskite, is a promising material for applications as electrolyte in solid oxide fuel cells[1-5]. BZO exhibits good proton conductivity at high temperatures[6,7]. In general, materials that exhibit high ionic conductivity tend to be less stable at high temperature[1,2]. However, BZO shows a good chemical and mechanical stability at these temperatures[8,9]. This may be explained by the relatively low diffusivity of the O vacancy at high temperatures[10-15]. However, the available experimental diffusion data[1,16-18] are not yet fully understood. Therefore, the present work is focused on first-principles Density-Functional-Theory (DFT) calculations of the free migration energy and the diffusivity of the doubly positively charged O vacancy in BZO. Most previous DFT studies on BZO were limited to ground state properties. However, recent investigations clearly demonstrate the inadequacy of using ground state data to explain finite temperature effects and emphasize the importance of considering phonon excitations. Bjørheim et al.[19] performed DFT calculations to understand the effect of phonon excitations on the free formation energy of neutral as well as doubly positively charged O vacancies and showed that these contributions strongly affect the relative stability of both vacancy types. For BZO Bjørheim et al.[20] also showed the significance of phonon contribution to surface segregation enthalpies and entropies of the doubly positively charged O vacancy and the hydroxide ion (OH$^-$). Hydrogen diffusion was studied by Sundell et al.[21] using DFT calculations and it was found that zero point vibrations effectively reduces activation enthalpies and make the prefactors similar for translation and rotation motion of H atom. While the effect of phonon excitations on the formation of the O vacancy has been studied[19], to the best of our knowledge, their influence on O vacancy migration has not been investigated yet. Therefore, in this work we determine the phonon contribution to free migration energy of the doubly positively charged O vacancy and compute the corresponding diffusivity. Further, we compare the calculated diffusivity with available experimental data.

The Vienna ab-initio simulation package (VASP) [22, 23] was used to perform DFT calculations. Plane wave basis sets and pseudopotentials generated within the projected augmented wave (PAW) approach were employed. In all calculations the generalized gradient approximation GGA-RPBE[24] was used to describe the exchange and correlation effects. The Brillouin zone sampling was performed using the Monkhorst-Pack scheme[25]. In all cases a plane wave cutoff of 500 eV was used. Calculations on the ground state energetics of BZO without and with a O vacancy were performed for a supercell with 40 lattice sites and Brillouin zone sampling of 4x4x4 k points as well as for a cell with 135 sites and 2x2x2 k points. After introduction of the defect into the supercell the positions of atoms as well as the volume and shape of the supercell were relaxed so that the total stress/pressure on the supercell became zero. Special care was taken to ensure a high precision of the results. This was achieved by performing several calculations to obtain the most suitable parameters for the iterations carried out in VASP. There are two important criteria to be considered: (i) First, zero pressure calculations are carried out until the residual force acting on any atom falls below a threshold of $10^{-6}$eV/Å, and (ii) at each step of ionic relaxation the relaxation of the electronic degrees of freedom is performed until the total energy change falls below another threshold of $10^{-8}$eV. To model doubly charged O vacancy, two electrons were removed from the overall supercell through charge compensation by a homogeneous jellium background charge.

The energy barrier for vacancy migration in BZO was evaluated using the Nudged Elastic Band method[26, 27], by considering the exchange of the O atom with the nearest neighbor vacancy. The states with the initial and final positions of the vacancy were connected by a number of system images that were constructed along the reaction path. In order to find an exact saddle point with a minimum number of images the climbing image method was employed as implemented in the "vtsttools"[28].

In order to determine the contribution of phonon excitations to the free energy, the vibrational frequencies of the corresponding supercell were calculated using the frozen phonon approach[29,30] and the harmonic approximation. In the present work the dynamical matrix with the force derivatives was determined using the ground-state atomic positions obtained under zero pressure conditions. The diagonalization of the dynamical matrix yields all vibrational frequencies $\omega_i$ of the supercell. Within the harmonic approximation the vibrational free energy $F^{vib}(T)$ of a system of $N$ atoms is given by

$$F^{vib}(T) = \sum_{i=1}^{3N-3} \left[ \frac{1}{2}\hbar\omega_i + k_B T \ln\left(1 - e^{-\frac{\hbar\omega_i}{k_B T}}\right) \right] \quad (1)$$

The vibrational contribution to the free migration energy of the doubly positively charged O vacancy is determined by the difference between the vibrational free energy at the saddle point and at the equilibrium state of the vacancy.

$$\Delta F_{mig}^{vib}(T) = F_{SP}^{vib}(T) - F_0^{vib}(T) \quad (2)$$

The quantities on the right-hand side of Eq. (2) are determined using Eq. (1), with the exception that in the first term, the sum is over $3N-4$ vibrational degrees of freedom. Within the harmonic transition state theory[31-33] the diffusivity of the O vacancy in the cubic BZO lattice is obtained by

$$D(T) = a^2 \frac{k_B T}{h} \exp\left(-\frac{\Delta F_{mig}^{vib}(T)}{k_B T}\right) \exp\left(-\frac{E_{mig}}{k_B T}\right) \tag{3}$$

Where, $a$ is the lattice parameter of BZO and $E_{mig}$ is the migration barrier of the O vacancy in the ground state.

Table I shows calculated data for bulk BZO and for BZO with the doubly positively charged O vacancy, in the equilibrium and the saddle configuration, in comparison with experimental and theoretical data from literature. The present results are in good agreement with reported experimental and theoretical data. The calculated lattice constant of BZO is 4.269Å which is slightly higher than the experimental lattice parameter[34] and the lattice parameter obtained with PBE functional[19]. For the O vacancy on equilibrium position the supercell shrinks as a consequence of the effective decrease of the bond distances. Since the O vacancy is doubly positively charged, there is significant local lattice distortion due to electrostatic interaction between the vacancy and the surrounding ions. The local lattice distortion is even higher in saddle configuration due to significant decrease of Ba-O and Zr-O bond distances (Table 1).

In BZO, the O vacancy moves to one of its first neighbour site, i.e. along the edge of the $ZrO_6$ octahedron. Figure 1 shows the O vacancy migration barrier in the ground state along with the migration path of the corresponding O ion. The O ion follows a curved trajectoryin contrast to a previous prediction of a linear path[35]. The maximum deviation from the linear path is 0.04 Å. The calculated ground-state migration barrier of 0.73 eV is in good agreement with the reported experimental result of 0.71 eV[18] and theoretical results of 0.65 eV[35] and 0.69 eV[36].

Figure 2 shows the relative deviation of the vibrational frequencies of the supercell with the O vacancy from the corresponding frequencies of the perfect BZO supercell. Results for the equilibrium as well as the saddle point configuration of the vacancy are depicted. The presence of the vacancy causes a decrease of the supercell volume and alters the equilibrium atomic positions of the ions. This leads to changes in the atomic force constants and, consequently, to a modification of the vibrational frequencies. For both vacancy configurations the majority of the data points are in the positive region, which indicates an overall shift to higher frequencies. This trend is illustrated by the straight lines which were obtained by a linear fit to the corresponding data points.

Figure 3 shows the difference between the vibrational free energy of the O vacancy in the saddle point and equilibrium configurations according to Eq.(2). At low and high temperatures this quantity shows a non-linear and a linear increase, respectively. At about 1000 K which is the normal operating temperature of Solid Oxide Fuel Cell (SOFC), the free migration energy of the O vacancy is 0.35 eV (or 50%) higher than its ground-state value. Such a large increase can be attributed to the blue shift of the vibrational frequencies obtained for the saddle point is higher than that of the respective frequencies in the equilibrium configuration (cf. Figure 2). As a result, the vibrational free energy at the saddle point increases faster with temperature than the vibrational free energy of the equilibrium configuration. The data obtained for the free migration energy of the O vacancy have direct bearing on the corresponding diffusivity. Figure 4 shows diffusivities calculated without and with considering the contributions of phonon excitations to the free migration energy of the O vacancy, together with data from measurements[1, 16-18]. The experimental results[1, 16-18] show large scatter due to different dopants in BZO, synthesis conditions, etc. However, these data may be used for comparison with our theoretical results since dopants should not have any direct bearing on the O vacancy migration. The diffusivity of the O vacancy calculated by taking into account the phonon excitations is in much better agreement with experimental data than the diffusivity obtained without contributions from phonons. At 1000 K, phonon contribution to free migration energy of O vacancy decreases its diffusivity by more than one order of magnitude.

In conclusion, the relatively low diffusivity of the doubly positively charged O vacancy in BZO can be explained by the considerable contribution of phonon excitations to the free migration energy. At ground state the free migration energy is 0.73 eV but at 1000 K it is 50% higher (1.08 eV). Present results reveal that the vibrational contribution to the free migration energy of the O vacancy is the main cause for the high-temperature stability of BZO.

Figure captions

Fig 1 (a) Migration barrier of the doubly positively charged oxygen vacancy in BZO along the reaction coordinate. (b) Two dimensional representation of the migration path of the corresponding O atom along the $ZrO_6$ octahedron edges. Green and red spheres represent Ba and O atoms respectively. Blue spheres represent migrating O atom. Zr atom is not shown for clarity.

Fig 2 Relative deviation of the vibrational frequencies calculated at the equilibrium and saddle configuration of the supercell with O vacancy from the vibrational frequencies of the bulk BZO supercell. The black and red symbols correspond to equilibrium and saddle configurations respectively. The straight lines were obtained by a linear fit to the data points and shall illustrate the trends.

Fig. 3 Difference between the vibrational free energy of doubly charged O vacancy in equilibrium configuration and the saddle point configuration. The inset shows the effective migration barrier at 1000 K (red circle) together with the ground state data from Fig. 1 (a).

Fig. 4 Diffusivity of the doubly positively charged O vacancy calculated with (blue solid line) and without (green solid line) contributions from phonon excitations, in comparison with experimental data (dotted lines - a (K.D. Kreuer, Ref. 1), b (H. G. Bohn et al., Ref.16), c (S. Tao et al., Ref. 17), and d (I. Ahmed et al., Ref. 18))

Table I: Properties of BZO without and with O vacancy; a,b,c: lattice constant (Å), d: bond lengths (Å), $\Delta a/a_0$, $\Delta b/b_0$, $\Delta c/c_0$: relative change of the lattice constant, $\Delta V/V_0$: relative volume relaxation, $\Delta d/d_0$: relative change of the bond length.

|  | Bulk | | | Oxygen vacancy (equilibrium position) | Oxygen vacancy (saddle point) |
| --- | --- | --- | --- | --- | --- |
|  | This Work | RPBE[19] | Expt[34] |  |  |
| a | 4.269 ($a_0$) | 4.269 | 4.192 | 4.218 | 4.218 |
| b | 4.269 ($b_0$) | 4.269 | 4.192 | 4.245 | 4.231 |
| c | 4.269 ($c_0$) | 4.269 | 4.192 | 4.218 | 4.231 |
| d(Ba-O) | 3.018 ($d_0$) | 3.019 | 2.969 | 2.918 | 2.540 |
| d(Zr-O) | 2.134 ($d_0$) | 2.135 | 2.099 | 2.079 | 1.924 |
| d(O-O) | 3.018 ($d_0$) | 3.019 | - | 2.907 | 2.859 |
| $\Delta a/a_0$ (%) |  |  |  | -1.194 | -1.194 |
| $\Delta b/b_0$ (%) |  |  |  | -0.562 | -0.890 |
| $\Delta c/c_0$ (%) |  |  |  | -1.194 | -0.890 |
| $\Delta V/V_0$ (%) |  |  |  | -2.916 | -2.902 |
| $\Delta d$(Ba-O)/$d_0$(Ba-O) (%) |  |  |  | -3.313 | -15.838 |
| $\Delta d$(Zr-O)/$d_0$(Zr-O) (%) |  |  |  | -2.577 | -9.841 |
| $\Delta d$(O-O)/$d_0$(O-O) (%) |  |  |  | -3.678 | -5.268 |

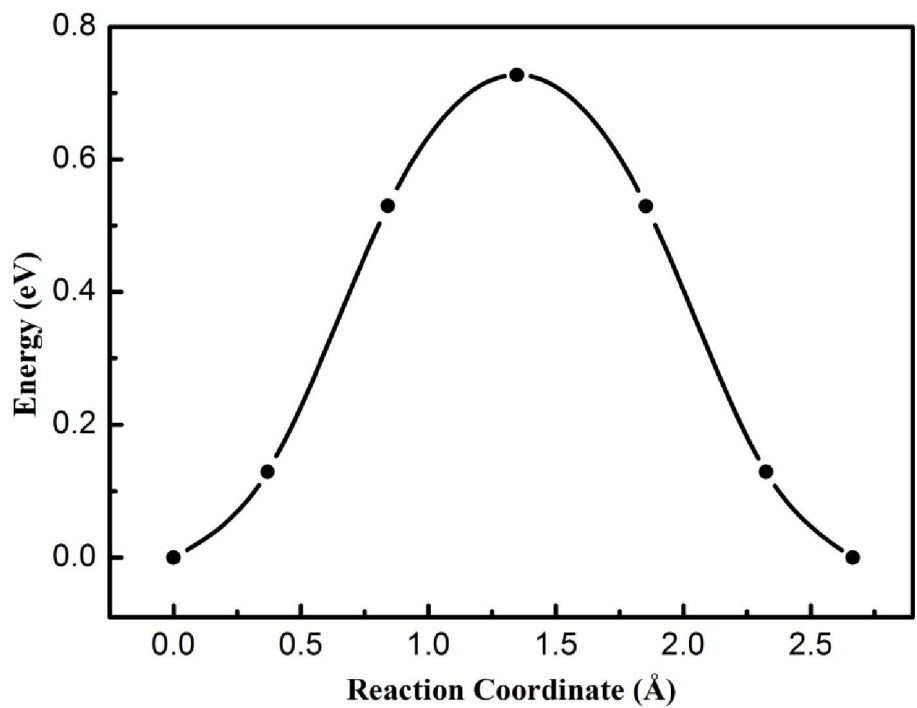

(a)

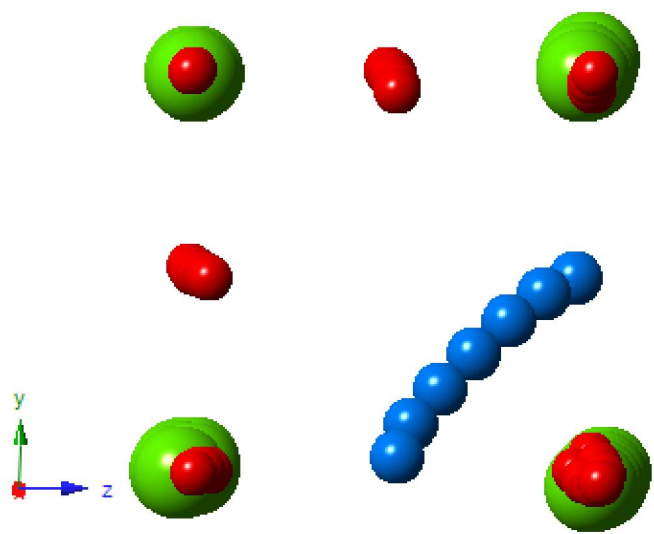

(b)

Figure 1

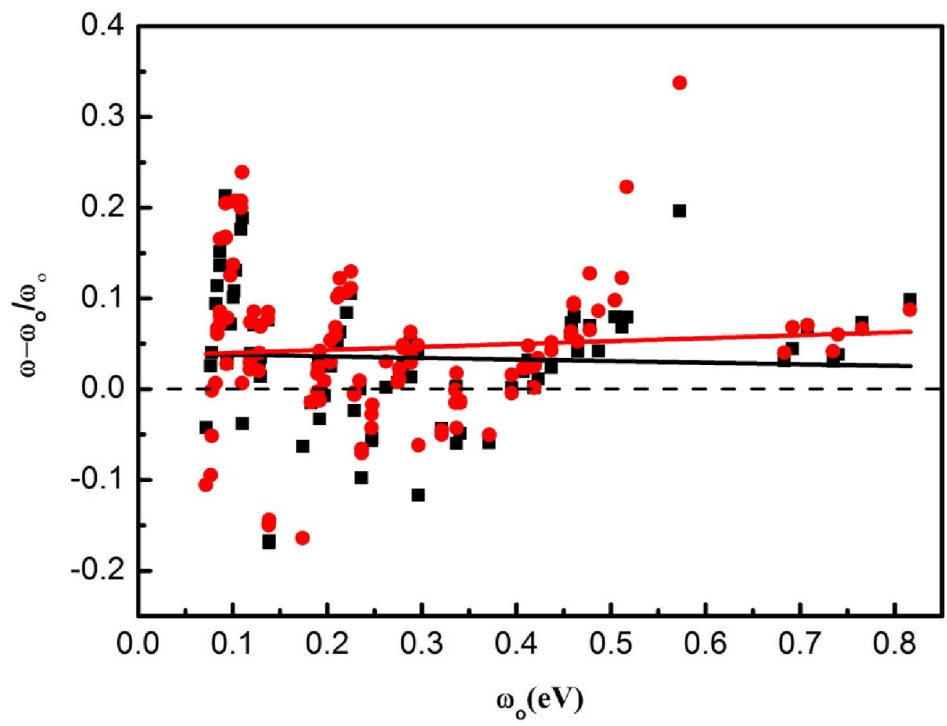

Figure 2

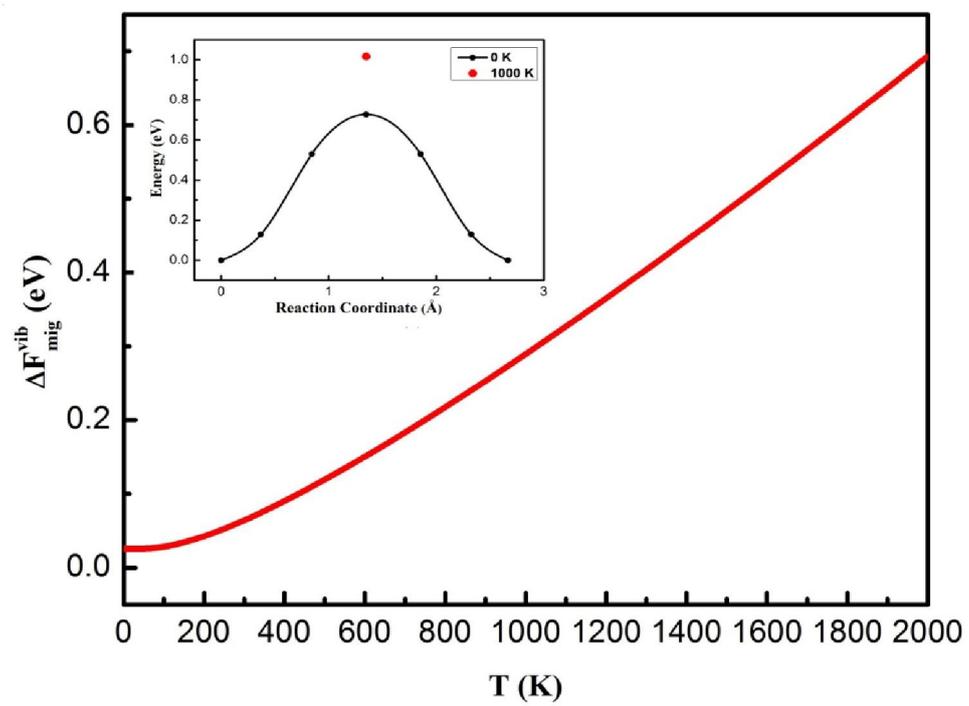

Figure 3

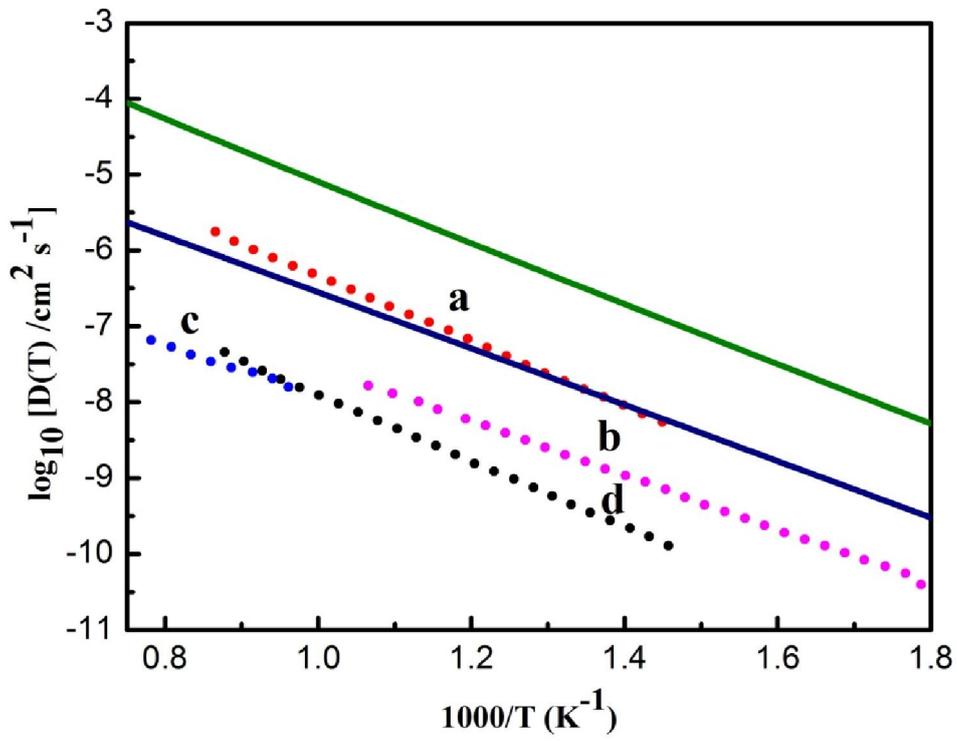

Figure 4